\documentclass[doublecol]{epl2} 

\title{Spontaneous symmetry breaking in Quantum Finance}
\shorttitle{Nambu-Goldstone theorem in Quantum Finance} 

\author{Ivan Arraut \inst{1} \and Alan Au\inst{1} \and Alan Ching-biu Tse \inst{2}}
\shortauthor{Ivan Arraut \etal}

\institute{                    
  \inst{1} The Open University of Hong Kong,\\
  30 Good Shepherd Street, Homantin, Kowloon\\
  \inst{2}  Department of Marketing, The Chinese University of Hong Kong, \\
  Cheng Yu Tung Building 12 Chak Cheung Street Shatin, N.T., Hong Kong
}
   
\pacs{}{Interdisciplinary applications of physics}
\pacs{89.65.Gh}{Economics, business, and financial markets}
\pacs{11.30.Qc}{Spontaneous and radiative symmetry breaking}

\abstract{We analyze the phenomena of spontaneous symmetry breaking in Quantum Finance by using as a starting point the Black-Scholes (BS) and the Merton-Garman (MG) equations expressed in the Hamiltonian form. In this scenario the martingale condition (state) corresponds to the vacuum state which becomes degenerate when the symmetry of the system is spontaneously broken. We then analyze the broken symmetries of the system and we interpret from the perspective of Financial markets the possible appearance of the Nambu-Goldstone bosons.}

\begin{document}

\maketitle

\section{Introduction}

Ordinarily the description of the dynamic of a physical system is done by analyzing its Hamiltonian (Lagrangian). This Mathematical object respects some symmetries, corresponding to transformations of some physical quantities such that the Hamiltonian (Lagrangian) remains unchanged up to total derivative terms, which are irrelevant for describing the dynamic of the system \cite{0}. In standard situations, the physical system has a single ground state which also respects the same symmetries of the Hamiltonian. There are special situations however, where the vacuum state violates some symmetries of the Hamiltonian \cite{1}. In the same situations, a multiplicity of ground states emerges. These circumstances correspond to the phenomena of spontaneous symmetry breaking \cite{2}. The spontaneous symmetry breaking phenomena is able to explain a significant number of observations, including superconductivity, ferromagnetism and the electroweak theory \cite{0, 1, 2}. Although some symmetry breaking approaches were analyzed in \cite{Referee} by using Financial equations, the generic phenomena of spontaneous symmetry breaking has not yet been explored in the scenario of Quantum finance. What we know about Quantum finance is that some financial equations such as the Black-Scholes (BS) \cite{9} and the Merton-Garman (MG) \cite{Op3, Merton2}, can be expressed in a Hamiltonian (eigenvalue) form after doing some transformation of variables \cite{B, 3, Non-Her, 4}. These equations then take the same structure of the Schr\"odinger equation in Quantum Mechanics \cite{Schro}. This allows us to use well-known techniques of Quantum Mechanics for doing some analysis \cite{3, Non-Her, 4}. Considering this point of view, the martingale condition is equivalent to the ground state of the system and it will be only modified if the information in the system flows or changes. In this letter we show that besides this observation, a series of interesting results emerge, they are: 1). The symmetry under changes of prices is spontaneously broken in these financial equations. This comes out to be connected with the flow of information through the boundaries of the system (market). 2). For the MG equation, the symmetry under changes in volatility is also spontaneously broken after extending the martingale condition in order to include the stochastic volatility as a variable. 3). Finally, we find a couple of interesting relations between volatility and the price of an option. One of them comes out when we consider the standard MG scenario where the symmetries under changes in the prices and volatility are both spontaneously broken. Another relation between the same variables occurs when we consider symmetry breaking patterns such that the price of the Option is maintained fixed. These kind of symmetry breaking patterns can be obtained after including some standard potential terms to the BS and to the MG equations. The inclusion of these potential terms must satisfy the martingale condition, which then forces some natural constraint between the new parameters entering the system.  

\section{The Martingale condition as a vacuum state}

The fundamental theorem of finance establishes the existence of a martingale measure as far as the market is complete and the condition of no-arbitrage holds. The martingale condition corresponds to a risk-neutral measure such that the conditional probability of the discounted value of an equity at time $t\neq0$ is just equal to its present value at $t=0$ \cite{B, M}. This can be expressed Mathematically as

\begin{equation}
S(0)=E[e^{-\int_0^tr(t')dt'}S(t)\vert S(0)].    
\end{equation}
Here $S(t)$ is just the value (price) taken by the equity and $r$ is short-term risk-free interest rate. In standard situations, the martingale condition is unique. Considering a standard change of variable $S=e^x$, we can formulate the the martingale condition as

\begin{equation}
S(x)=\int_{-\infty}^\infty dx'<x\vert e^{-(t_*-t)\hat{H}}\vert x'>S(x').    
\end{equation}
If we use the complete base condition $\int_{-\infty}^\infty dx\vert x><x\vert=\hat{I}$, with $\hat{I}$ being the identity, and in addition $S(x)=<x\vert S>$, then we obtain

\begin{equation}
\vert S>=e^{-(t_*-t)\hat{H}}\vert S>.    
\end{equation}
The same condition can be expressed as 

\begin{equation}   \label{vacuum}
\hat{H}\vert S>=0.    
\end{equation}
This condition can be also verified by applying the explicit form of the financial Hamiltonian over the state $\vert S>$. The result (\ref{vacuum}) corresponds to the ground state of the system or equivalently, to the vacuum state. Eq. (\ref{vacuum}) is the martingale condition expressed from the perspective of Quantum finance \cite{B}.

\section{Broken symmetries in the financial equations}

If there exist a symmetry of a financial Hamiltonian defined as $[\hat{H}, \hat{A}]=0$ and if in addition $\hat{A}\vert S>\neq0$, then we say that the symmetry is spontaneously broken \cite{1, 2}. This means that the vacuum (martingale) state does not obey the same symmetries of the Hamiltonian defining the dynamic of the system. Here we will provide a few examples. We can start with the BS equation expressed in the Hamiltonian form \cite{B}

\begin{equation}   \label{Hamil}
\hat{H}_{BS}C(x, t)=EC(x, t),    
\end{equation}

\begin{equation}   \label{BSHamiltonian}
\hat{H}_{BS}=-\frac{\sigma^2}{2}\hat{p}^2+\left(\frac{1}{2}\sigma^2-r\right)\hat{p}+r.
\end{equation}
Here $<x\vert\hat{p}\vert C>=\partial C(x, t)/\partial x$. It can be easily proved that if we replace the option $C(x, t)$ in eq. (\ref{Hamil}) by the security $S(x, t)=e^x$, then we obtain the vacuum condition (\ref{vacuum}). Note however that although the operator defined as $\hat{p}C(x, t)=\partial C(x, t)/\partial x$ is a conserved quantity (symmetry of the Hamiltonian), the same operator is not a generator for the symmetry of the ground (martingale) state since

\begin{equation}   \label{Spo}
\hat{p}\vert S>\neq0.    
\end{equation}
Then this symmetry is spontaneously broken and it corresponds to the invariance of the Hamiltonian under the changes of the prices of the options. Then from this perspective we have a degenerate vacuum. This should not be a surprise since this simply means that any value taken by the security $S(x, t)=e^x$ represents a martingale as far as the market is complete and there is no chance of arbitrage. This in addition implies that the market can reach the equilibrium for any value of $S(x, t)$. Equivalently, for any ket $\vert S>$, we have potentially a martingale condition if the market is in equilibrium. Note that if $\hat{p}$ is extended to the complex plane such that it becomes $\hat{p}\to i\hat{p}$, then  symmetry spontaneously broken would correspond to rotations on the complex plane of the form $U=e^{-i\hat{p}\alpha}$. The previously mentioned results and conclusions can be done for the MG Hamiltonian defined as \cite{B}

\begin{eqnarray}  \label{MGHamilton}
\hat{H}_{MG}=-\frac{e^y}{2}\hat{p}_x^2-\left(r-\frac{e^y}{2}\right)\hat{p}_x\nonumber\\
-\left(\lambda e^{-y}+\mu-\frac{\zeta^2}{2}e^{2y(\alpha-1)}\right)
\hat{p}_y\nonumber\\
-\rho\zeta e^{y(\alpha-1/2)}\hat{p}_x\hat{p}_y-\zeta^2 e^{2y(\alpha-1)}\hat{p}_y^2+r.   
\end{eqnarray}
Here again we have defined the generator of the changes in prices as $<x, y\vert\hat{p}_x\vert C>=\partial C(x, y, t)/\partial x$ and the generator under changes of volatility as $<x, y\vert\hat{p}_y\vert C>=\partial C(x, y, t)/\partial y$. Note that here again if we apply $\hat{H}_{MG}\vert S>$, we will obtain trivially the result (\ref{vacuum}), taking into account that $S(x, t)=<x\vert S>$ only depends on the variable $x$ and then the derivatives with respect to $y$ vanish trivially. Here the variable $y$ is related to the stochastic volatility defined as $\sigma^2=V=e^y$ \cite{B}. If we define $\hat{p}_xC(x, t)=\partial C(x, t)/\partial x$ as the "momentum" associated to the Stock prices and $\hat{p}_yC(x, t)=\partial C(x, t)/\partial y$ as the "momentum" related to the volatility, then we conclude that for the standard MG situation, only the momentum related to the prices is spontaneously broken. 

\section{Spontaneous symmetry breaking: Symmetries under changes of prices}

From the result (\ref{Spo}), we can derive some general expressions if we define the martingale state in agreement with $<x\vert S>=e^x=\sum_n\phi^{n}$, such that $<x\vert S>=S(x, t)$. The effect of the broken generator $\hat{p}$ over the field $\phi$ is to map it toward another field $\bar{\phi}$ with different components. In the same way, applying $\hat{p}$ over $\bar{\phi}$ appropriately, maps it back to the original field $\phi$. Without loss of generality, we can select $\bar{\phi}$ to satisfy $<S\vert\bar{\phi}\vert S>=0$. Taking into account these definitions, we can evaluate the commutator

\begin{eqnarray}   \label{superimportant}
<S\vert[\hat{p}, \bar{\phi}]\vert S>=\int_{-\infty}^\infty dxS(x, t)\frac{\partial}{\partial x}\bar{\phi}(x, t)\neq0,
\end{eqnarray}
which is obtained from the expression 

\begin{eqnarray}
\int_{-\infty}^\infty dx<S\vert \hat{p}\vert x><x\vert\bar{\phi}\vert S>-\nonumber\\
\int_{-\infty}^\infty dx<S\vert \bar{\phi}\vert x><x\vert\hat{p}\vert S>, 
\end{eqnarray}
and taking into account the completeness relation $\int_{-\infty}^\infty dx\vert x><x\vert=\hat{I}$, together with the standard expression $<x\vert\hat{p}\vert S>=\partial S/\partial x$. Additionally, we define $<x\vert\bar{\phi}\vert S>=\bar{\phi}(x, t)$. Note that the previous result is general and it does not depend on how we define the field $\bar{\phi}$. Inside the previous results, we can demonstrate easily that $<S\vert \phi\vert S>\neq0$ because $<S\vert\sum_n\phi^{n}\vert S>=e^x$, which never vanishes unless $x\to-\infty$. Note however that here we have fixed $\bar{\phi}$ as a Quantum field satisfying $<S\vert\bar{\phi}\vert S>=0$, which precisely corresponds to $x\to-\infty$. This is the Nambu-Goldstone field. Additionally, we can demonstrate that the effect of $\hat{p}$ is to move the system from one vacuum to another one, let's say from the vacuum defined by $\bar{\phi}$ toward the vacuum defined by $\phi$. In this way we have the definition of order parameter as

\begin{equation}   \label{prev}
<S\vert[\hat{p}, \bar{\phi}]\vert S>=<S\vert\phi\vert S>\neq0,
\end{equation}
which is valid if 

\begin{equation}   \label{spiderman}
\hat{p}\bar{\phi}(x)=\frac{\partial\bar{\phi}(x)}{\partial x}=\phi(x),
\end{equation}
taking into account that 

\begin{eqnarray}
<S\vert \phi\vert S>=\int_{-\infty}^\infty dx<S\vert x><x\vert \phi\vert S>=\nonumber\\
\int_{-\infty}^\infty dx S(x, t)\phi(x), 
\end{eqnarray}
in general. The consistency of this expression with the results (\ref{superimportant}) and (\ref{prev}), suggests the validity of the condition (\ref{spiderman}). Then what the broken symmetry generator $\hat{p}$ does is to map one vacuum toward another one, this is typical from spontaneous symmetry breaking. Another way to visualize these results is by defining the potential of the Hamiltonian (\ref{BSHamiltonian}) as follows

\begin{equation}   \label{pot}
\hat{V}=\left(\frac{1}{2}\sigma^2-r\right)\hat{p}+r.    
\end{equation}
Around the minimal, the state $C(x, t)$ approaches to $S(x, t)=e^x$, which is the martingale condition. Considering the expansion $S(x, t)=e^x=\sum_{n=0}^\infty (x)^n/n!=\sum_{n=0}^\infty \phi^n(x, t)$, we can define the potential (\ref{pot}) as a function of $S(x, t)$ taken as a field as

\begin{eqnarray}
V(S)=<x\vert\hat{V}\vert S>= \sum_n\left(\frac{1}{2}\sigma^2-r\right)n\phi^{n-1}(x)+\nonumber\\
r\phi^n(x). 
\end{eqnarray}
Here we have used the definition of $\phi^n(x, t)$ and the fact that $\partial\phi^n/\partial x=n\phi^{n-1}$. This result is obtained if we consider $\partial S/\partial x=e^x=\sum_{n=0}^\infty (x)^{n-1}/(n-1)!=\sum_{n=0}^\infty n\phi^{n-1}(x, t)$. This means that $\sum_n\partial\phi^n(x, t)/\partial x=\sum_n\phi^n(x, t)=\sum_n n\phi^{n-1}(x, t)$. However, since we have to compare terms in the expansion order by order, we have to pick up any order in the expansion of $n$ for doing the corresponding calculations. In this way, without loss of generality, we can just focus on the second order terms in the expansion ($n=2$) as follows   

\begin{equation}   \label{Thisone}
V(S)=<x\vert \hat{V}\vert S>\approx 2\left(\frac{1}{2}\sigma^2-r\right)\phi(x)+r\phi(x)^2.    
\end{equation}
What is really important is the relative exponent between the derivative and the non-derivative term. This relation will not change, no matter what order in the expansion we decide to compute. Then for simplicity we can consider eq. (\ref{Thisone}) as the appropriate potential in order to analyze the vacuum conditions of the system. In this way, we can express the minimal of the potential as a function of the free-parameters of the theory, namely $\sigma$ and $r$. Note that the potential (\ref{Thisone}) contains a term which comes out from the derivative-term in eq. (\ref{pot}) ($\hat{p}S(x, t)=\partial S(x, t)/\partial x$), which is non-Hermitian. Indeed, this term is related to the flow of information through the boundaries of the system \cite{Non-Her}. The minimal for the potential (\ref{Thisone}) occurs when $\frac{\partial V(S)}{\partial\phi}=0$. Solving for $\phi(x)$, then we find

\begin{equation}   \label{nonunique}
\phi_{vac}=1-\frac{\sigma^2}{2r}.     
\end{equation}
This vacuum condition will then be fixed depending on the interest rate and volatility. Note that when $r=\sigma^2/2$, then the vacuum is trivial since $\phi_{vac}=0$. This naturally corresponds to a trivial value of the security $S(x, t)$. The same relation guarantees the no-flow of information through the boundaries of the system. Here we take $\sigma^2\leq 2r$ conventionally for avoiding the theory to be unstable. Then the largest possible value for the volatility is constrained by the interest rate $r$. Note additionally that when $r>>\sigma^2$, the vacuum converges to a constant value. The relation (\ref{nonunique}) is non-unique in the sense that we can define the vacuum arbitrarily in principle. However, what is really important here is to take into account that the vacuum depends on the relation between the fundamental constants of the system. We can repeat the same arguments for the MG Hamiltonian (\ref{MGHamilton}), finding then the same relation (\ref{nonunique}), but with $\sigma^2=e^y$. 

\section{Extended martingale condition including the stochastic volatility for the Merton Garman case}

In the standard case of the MG equation, the martingale condition is taken to be independent of the stochastic volatility, defined through the variable $y$ in eq. (\ref{MGHamilton}). However, we can still define a different vacuum condition with some dependence on $y$ as follows

\begin{equation}   \label{newmartin}
\hat{H}_{MG}e^{x+y}=\hat{H}_{MG}S(x, y, t)=0.    
\end{equation}
The Hamiltonian (\ref{MGHamilton}) annihilates the vacuum as far as the following condition is satisfied

\begin{equation}   \label{corona4}
\lambda+e^y\left(\mu+\frac{\zeta^2}{2}e^{2y(\alpha-1)}+\rho\zeta e^{y(\alpha-1/2)}\right)=0.
\end{equation}
This condition is necessary, except for the trivial case where $e^x=0$. We can define the potential for the Hamiltonian (\ref{MGHamilton}) as 

\begin{eqnarray}   \label{happotential}
\hat{V}(x, y)= -\left(r-\frac{e^y}{2}\right)\hat{p}_x\nonumber\\  
-\left(\lambda e^{-y}+\mu-\frac{\zeta^2}{2}e^{2y(\alpha-1)}\right)\hat{p}_y+r.
\end{eqnarray}
In the neighborhood of the minimal defined by the condition (\ref{newmartin}), we can obtain interesting results. Considering that $<x, y\vert S>=S(x, y, t)=e^{x+y}=\sum_{n=0}^\infty(x+y)^n/n!=\sum_{n=0}^\infty\phi^n_x\phi^n_y$. Additionally, considering that $\partial S(x, y, t)/\partial x=\partial S(x, y, t)/\partial y=\sum_n(x+y)^{n-1}/(n-1)!=e^{x+y}=\sum_nn\phi_x^{n-1}\phi_y^n=\sum_nn\phi_x^n\phi_y^{n-1}$, we can easily evaluate the potential term in eq. (\ref{happotential}) as  

\begin{eqnarray}
<x, y\vert\hat{V}\vert S>=-\sum_nn\left(r-\frac{e^y}{2}\right)\phi_x^{n-1}\phi_y\nonumber\\
-\sum_nn\left(\lambda e^{-y}+\mu-\frac{\zeta^2}{2}e^{2y(\alpha-1)}\right)\phi_x\phi_y^{n-1}+r\sum_n\phi_x^n\phi_y^n.
\end{eqnarray}
Here again without loss of generality we can focus only on the second order terms in the expansion. Any other order in the expansion will provide the same results. Then our simplified potential term is

\begin{eqnarray}
V(S)=-2\left(r-\frac{e^y}{2}\right)\phi_x\phi_y^2\nonumber\\
-2\left(\lambda e^{-y}+\mu-\frac{\zeta^2}{2}e^{2y(\alpha-1)}\right)\phi_x^2\phi_y+r\phi_x^2\phi_y^2.    
\end{eqnarray}
The minimal for this potential occurs when $\partial V/\partial\phi_x=\partial V/\partial\phi_y=0$. Solving these two equations, then we get the relation

\begin{equation}   \label{trivacc}
\phi_{y vac}=\left(\frac{\lambda e^{-y}+\mu-\frac{\zeta^2}{2}e^{2y(\alpha-1)}}{r-\frac{e^y}{2}}\right)\phi_{x vac},    
\end{equation}
with the full vacuum defined by $S(x, y)=\phi_{x vac}\phi_{y vac}$. Note that the vacuum state $S(x, t)$ becomes trivial when either of the non-Hermitian terms in the Hamiltonian (\ref{MGHamilton}) vanishes. This happens when 

\begin{equation}   \label{volalala}
r=\frac{e^y}{2},\;\;\;or\;\;\;\lambda e^{-y}+\mu-\frac{\zeta^2}{2}e^{2y(\alpha-1)}=0.    
\end{equation}
Any other combination of parameters avoiding these two conditions represents a non-trivial vacuum state. Note that the first condition in eq. (\ref{volalala}) represents the result $\phi_{xvac}=0$ and $\phi_{yvac}\neq0$. This means a trivial case in the prices of the Stock with arbitrary volatility. The second condition in eq. (\ref{volalala}) represents the situation $\phi_{xvac}\neq0$ and $\phi_{yvac}=0$, which represents a constant value of volatility for any price. In this way we can see that in general there is a connection between the definition of vacuum and the flow of information. A trivial vacuum represents a zero value for the security $S(x, y)$ and the fact that this happens when one of the non-Hermitian contributions vanishes supports this statement. The MG potential has in this case two flat directions representing the paths where the Nambu-Goldstone bosons should flow \cite{2}. Here of course we interpret the Nambu-Goldstone bosons as the fields connecting the different prices related to the martingale condition. In the same way, there are two broken symmetries corresponding to the broken generators defined as

\begin{equation}
\hat{p}_x\vert S>\neq0,\;\;\;\;\;\hat{p}_y\vert S>\neq0.    
\end{equation}
All the arguments discussed previously are valid, except that for this case we have two broken symmetries instead of only one as in the standard case. The result (\ref{trivacc}) is very useful at the moment of analyzing data in the financial market. After calibration, this expression is fixing the prices of the Stocks (represented by the variable $x$) in equilibrium with some specific values in the volatility of the market (represented by the variable $y$). Note the the variables $x$ and $y$ enter as a series expansion through the corresponding fields $\phi_x$ and $\phi_y$. The relation between price and volatility is then fixed by all the possible values taken by the free-parameters of the system, namely, $\lambda$, $\mu$, $\zeta$, $r$ and $\alpha$, which are originally connected to the dynamic of the stochastic volatility \cite{B}. A non-trivial value for $\zeta$ in particular is connected to the random white noise of the volatility \cite{B}.       

\section{Broken symmetries keeping the price of the security unchanged}

Now we will explain other possible broken symmetries besides those connecting different prices of the security $S$, which defined the vacuum (martingale) state. For the case of spontaneous symmetry breaking under the changes in the prices, it is not a trivial issue to notice that the vacuum is degenerate, we can visualize this if we extend the operator $\hat{p}$ to be Hermitian as $\hat{p}\to i\hat{p}$ such that we define the transformations related to changes in prices as $U=e^{-ipx}$, with $U$ being a unitary operator. In this case, it is evident that different prices (different values for $x$) related to the Martingale state would correspond to different phases (angles) which correspond to different definitions of vacuum. Evidently, the different vacuums contain different amount of information and because of this, there is a direct connection between the flow of information (non-Hermiticity of the Hamiltonian) and the different definitions of vacuum as we just illustrated recently. In this section we will consider additional symmetries which can be broken spontaneously due to external circumstances, but at the same time, they keep the prices unchanged. Note that the BS equation as well as the MG equation are idealizations of the market. In order to model real situations, in general we can include potential terms in the Hamiltonian. Here we will include some arbitrary potential, which as a field, will be a function of the prices of the option $C(x, t)$. For the MG as well as for the BS equation, we can define the potential term of the the system as

\begin{equation}   \label{Mix}
\hat{H}=\hat{H}_{BS, MG}(\partial/\partial x)+\hat{V}.    
\end{equation}
The first part corresponds to the portion of the Hamiltonian which only contains derivative terms. The potential will contain non-derivative terms. This special form of the potential is typical of the situations where the prices are not affected by the symmetry breaking transformations.
These transformations are equivalent to rotations which keep $x$ constant. Then the generator of the transformation is analogous to the angular momentum in Quantum Mechanics $\hat{L}_x$, pointing in the $x$-direction (no change in $x$). Note that since $[\hat{L}_x, \hat{p}_x]=0$ (in general $[\hat{L}_i, \hat{p}_j]=i\hbar\epsilon_{ijk}\hat{p}_k$, with $\epsilon_{ijk}$ representing the Levi-Civita symbol), then $\hat{L}_x$ is still a conserved quantity in the sense $[\hat{H}, \hat{L}_x]=0$. If we consider eq. (\ref{Mix}) as a field equation, after taking $C(x, t)$ as a field, then we have

\begin{equation}
<x, y\vert\hat{H}_{BS, MG}\vert C>=Kinetic\;\;terms +rC(x, t).
\end{equation}
Note that the potential term here is linear, its minimal would correspond to a condition of zero interest rate $r=0$. This trivial situation is not interesting for the cases where we analyze symmetry transformations with fixed prices. A more interesting situation would involve higher-order contributions. In order to keep the martingale condition fixed, we need to add at least two additional terms involving two additional free-parameters of the theory. Then we could express the modified BS and MG equations as 

\begin{eqnarray}   \label{Mix2}
<x, y\vert\hat{H}\vert C>\approx <x, y\vert\hat{H}_{BS, MG}\vert C>+\mu^2C^2(x, t)\nonumber\\
+\lambda C^4(x, t).    
\end{eqnarray}
In principle, any other combination would be possible. We could include for example a term of the form $C^3(x, t)$ instead of a term like $C^4(x, t)$ in eq. (\ref{Mix2}). However, the key points of the analysis will not change. The type of terms added to the Hamiltonian will depend on additional issues such as the type of symmetries which we would like the Hamiltonian to respect. Here as an example we will take our Hamiltonian as in eq. (\ref{Mix2}). We already know that the ground state condition is defined by a martingale state as the one defined in eq. (\ref{vacuum}). This conditions is satisfied by both, the MG as well as well as the BS equations. In order to keep this condition unchanged under the addition of terms, the new added higher-order contributions must satisfy

\begin{equation}
\hat{V}_{new}(S)=0=\mu^2S^2(x, t)+\lambda S^4(x, t).    
\end{equation}
This gives us the condition 

\begin{equation}   \label{fixed}
S=\pm\left(-\frac{\mu^2}{\lambda}\right)^{1/2},   
\end{equation}
for the vacuum state. This result is valid when $\lambda<0$. For positive values of $\lambda$, the martingale condition would be trivially $S(x, t)=0$. For the case of the MG equation, even considering the situations where we include the volatility as a variable as we did in eq. (\ref{newmartin}), we will still obtain the same fixed norm for the martingale as in eq. (\ref{fixed}). However, since the MG equation has two degrees of freedom \cite{B}, then eq. (\ref{fixed}) would fix the magnitude of the vacuum state but not its direction. In  other words, we would have the important relation

\begin{equation}   \label{fixed2}
e^x=\pm e^{-y}\left(-\frac{\mu^2}{\lambda}\right)^{1/2}.   
\end{equation}
Here we have used the result (\ref{newmartin}), together with eq. (\ref{fixed}), taking into account that $S(x, y, t)=e^{x+y}$. The result (\ref{fixed2}) illustrates a compromise between the variables $x$ and $y$ once a vacuum is fixed. This marks a fixed relation between volatility (through the variable $y$) and the price of the security (through the variable $x$). It is then clear that when the volatility grows, the price of the security decreases for a fixed vacuum condition. This is consistent with some observations in the market. Note that in this case, the behavior between price and volatility described by the vacuum condition is different with respect to the situation described inside the standard MG equation with an extended martingale condition and given in eq. (\ref{trivacc}). Then after calibration, we can see if the Option under analysis in the stock market follows a process where the amount of information changes as it is described by eq. (\ref{trivacc}) or if instead the equilibrium of the system follows a process with the compromise between variables defined in eq. (\ref{fixed2}). This will help us to identify different circumstances affecting the behavior in the market in general. Here evidently, although the Hamiltonian would be unchanged under the transformation $U=e^{-i\hat{L}_x\theta}$, due to the fact that the symmetry is spontaneously broken, we would have 

\begin{equation}
\hat{L}_xS(x, y, t)\neq0.    
\end{equation}
This means that 

\begin{equation}
U\vert S>=e^{-i\hat{L}_x\theta}\vert S>=\vert S'>\neq\vert S>.
\end{equation}
Then the action of $U$ in this case is to map one martingale condition to another one. Note that this situation is different to the ordinary BS or MG equations, in particular, different to the case where we analyzed the symmetries under changes of prices. Here for example, since the prices do not change, each vacuum (martingale state), has exactly the same amount of information. What really changes from one vacuum to the next is how you distribute such information. Before concluding, we have to remark something particular about the MG equation in relation with the symmetries generated by $\hat{L}_x$. Note that $\hat{L}_x$ is a conserved quantity, namely, $[\hat{H}_{MG}, \hat{L}_x]=0$ if the non-Hermitian term proportional to $\hat{p}_y$ in eq. (\ref{MGHamilton}) vanishes. This happens when 

\begin{equation}
\lambda e^{-y}+\mu-\frac{\zeta^2}{2}e^{2y(\alpha-1)}=0.    
\end{equation}
Finally, we can extend this analysis to another broken symmetry defined by the generator $\hat{L}_y$ which corresponds to a symmetry which keeps the volatility (not the prices) fixed. This would be analogous to a rotation generated by an angular momentum $\hat{L}_y$, pointing in the $y$-direction. The analysis will not be different to what has been explained before, except for the fact that $\hat{L}_y$ is a symmetry generator for the Hamiltonian $\hat{H}_{MG}$, only when the non-Hermitian term proportional to $\hat{p}_x$ in eq. (\ref{MGHamilton}) vanishes. This happens when

\begin{equation}
r=\frac{e^y}{2}.    
\end{equation}
Note that in this case again the same condition (\ref{fixed2}) will emerge. Then the conclusions do not change, except for the fact that the constraints related to the volatility are different.    

\section{Conclusions}

In this letter we have analyzed the concept of Spontaneous symmetry breaking in Quantum Finance, by using as a starting point two different Financial equations, namely, the BS equation and the MG equation. We demonstrated that another way to analyze these equations in their standard form, is by considering the concept of spontaneous symmetry breaking. When both equations are expressed as a Hamiltonian equation of the Schr\"odinger type, then it comes out that the martingale state is the ground (vacuum) state of the system. This ground state comes out to be degenerate, considering the fact that the symmetry under changes of prices is spontaneously broken. We then explained how to analyze this issue by expressing the BS and the MG equations as a field equations. Subsequently we extended the concept of martingale, for the MG equation, in order to include the changes in volatility in the definition of vacuum. When this is the case, not only the symmetries under changes of prices, but also the symmetry under changes of volatility is spontaneously broken. Interestingly, it came out from these situations that each vacuum has different information and that the martingale state becomes trivial when there is no flow of information through the boundaries of the system. This is the case if we take into account that the non-Hermiticity of the Hamiltonian is related to the flow of information in the system \cite{Non-Her}. Finally, we analyzed situations where the symmetry of the system is spontaneously broken but keeping the prices unchanged. This happens when the symmetry generator under analysis is analogous to the angular momentum pointing in either, the $x$-direction or the $y$-direction. Physically we interpret this case as a degenerate vacuum where the martingale states are represented by the same price but with different kind of information. By different kind of information we mean different events, in this case being able to reproduce the same effects in the stock market such that the prices under analysis are just the same. As an example, when two countries go to a war, this event has an effect in the prices of some stocks and then the system reaches some new equilibrium in prices based on this information. However, probably the same new equilibrium and effects might appear due to an completely unrelated event like an economical blocking (trade war for example) or anything else. Then instead of identifying two events affecting the Stock market as different, we can connect them by using the concept of spontaneous symmetry breaking proposed here. Note that we have to add additional non-derivative terms to the standard Hamiltonian in order to get the effects of having degenerate vacuums (multiplicity of martingales) with the same price. The new terms together must evidently obey the Martingale condition. Note that previously some authors considered some symmetry breaking mechanism in the MG case \cite{Referee}, however, the kind of symmetries as well as the methods developed were different to the case under study in the present paper. Finally, in a financial setting, the methods analyzed here can be used for improving some predictions in the market. In particular, inside the MG case, our symmetry breaking methods defined an interesting compromise between the price and volatility as can be observed in eq. (\ref{trivacc}). This equation suggests a proportionality relation between volatility represented through the field $\phi_y$ and the prices of the Stock, represented through the field $\phi_x$. Both fields are series expansions of the corresponding variables. The relation (\ref{trivacc}) suggests that in equilibrium, the values taken by the prices of a Stock are fixed to some specific values of volatility in the market; and the same values will mainly depend on the free-parameters of the system. A different set-up appears when we consider the situations where the prices of the Stocks do not change. In such a case, the extended MG case suggests an equilibrium condition defined in eq. (\ref{fixed2}), where the relation between prices and volatility changes, becoming in this way, an inversely proportional relation between the price and the volatility. The same relation in such situations would now depend surprisingly on only two free-parameters. These are just examples about how the important methods of symmetry breaking mechanisms can be applied to the market.   

\acknowledgments
{\bf Acknowledgments}\\
The authors also would like to thank the Institute of
International Business and Governance of the Open University of Hong Kong, partially funded by a grant from the
Research Grants Council of the Hong Kong Special Administrative Region, China (UGC/IDS16/17), for its support.

\end{document}